\documentclass[journal=jctc,manuscript=article]{achemso}
\pdfoutput=1
\usepackage[version=3]{mhchem} 
\usepackage{amsfonts}
\usepackage{physics}

\author{Gustavo J. R. Aroeira}
\author{Kyle Kairys}
\author{Raphael F. Ribeiro}
\email{raphael.ribeiro@emory.edu}
\affiliation[Emory University]
{Department of Chemistry and Cherry Emerson Center for Scientific Computation, Emory University, Atlanta, Georgia, United States of America}

\title{Theoretical Analysis of Exciton Wave Packet Dynamics in Polaritonic Wires}

\begin{document}

\begin{abstract}
We present a comprehensive study of exciton wave packet evolution in disordered lossless polaritonic wires. Our simulations reveal signatures of ballistic, diffusive, and subdiffusive exciton dynamics under strong light-matter coupling and identify the typical timescales associated with the transitions between these qualitatively distinct transport phenomena. We  determine optimal truncations of the molecular subsystem and radiation field required for generating reliable time-dependent data from computational simulations at affordable cost. The time evolution of the photonic part of the wave function reveals that many cavity modes contribute to the dynamics in a non-trivial fashion. Hence, a sizable number of photon modes is needed to describe exciton propagation with reasonable accuracy. We find and discuss an intriguingly common lack of dominance of the photon mode on resonance with the molecular system both in the presence and absence of disorder. We discuss the implications of our investigations to the development of theoretical models and analysis of experiments where coherent intermolecular energy transport and static disorder play an important role.
\end{abstract}

\section{Introduction}

\par Interaction between light and matter is enhanced within optical microcavities and plasmonic devices due to the confinement of the electromagnetic (EM) field to a small region of space. \cite{kavokin2017microcavities}. These structures have been used to design landscapes where the strong light-matter coupling regime achieved enables the emergence of light-matter hybrid states commonly denoted (cavity) polaritons. \cite{Lidzey2015StrongStructures, Ebbesen2016HybridPerspective}. The presence of these polaritonic states has been shown to modify energy transport,\cite{Coles2014Polariton-mediatedMicrocavity, Zhong2017EnergyMolecules, Xiang2020IntermolecularCoupling, Hagenmuller2017Cavity-EnhancedCharge, Hagenmuller2018Cavity-assistedDynamics, Garcia-Vidal2021ManipulatingFields,  Wang2021Polariton-assistedHeterojunctions, Guo2022BoostingSubstrate} conductivity and photoconductivity,\cite{Orgiu2015ConductivityField, Krainova2020PolaronRegime, Nagarajan2020ConductivityCoupling, Bhatt2021EnhancedCoupling, Liu2022PhotocurrentExcitonpolaritons} optical response,\cite{Ebbesen2016HybridPerspective, Dunkelberger2022Vibration-CavityDynamics} and chemical reactions\cite{Hirai2020RecentChemistry, Garcia-Vidal2021ManipulatingFields, Li2022MolecularCoupling, Dunkelberger2022Vibration-CavityDynamics, Galego2016SuppressingFields}. Hence, these devices have been not only objects of theoretical interest but also prospectus of new technology\cite{Imamoglu96, Sanvitto2016TheDevices, kavokin2017microcavities}.

Energy transfer mediated by microcavity polaritons was first verified with a binary J-aggregate mixture using photoluminescence \cite{Coles2014Polariton-mediatedMicrocavity} followed by femtosecond transient spectroscopy \cite{Zhong2016Non-RadiativeStates}. Even when donor and acceptor molecules were spatially separated, energy transfer was observed, ruling out the possibility of dipole-dipole energy transfer \cite{Zhong2017EnergyMolecules}. Transport distances of approximately 20 $\mu$m have been reported for inorganic quantum well polaritons with low photon content in a two-dimensional planar cavity \cite{Myers2018Polariton-enhancedTransport}. In contrast, polaritons emergent from strong coupling of Bloch surface waves and organic materials showed $\mathcal{O}(10^2 )$ $\mu$m propagation lengths \cite{Lerario2017High-speedPolaritons, Hou2020UltralongRangePropagation}. Corresponding group velocities of over 120 $\mu$m ps$^{-1}$ were also deduced from dispersion relations \cite{Lerario2017High-speedPolaritons}. Direct measurements of exciton-polariton pulse widths have recently revealed growth rates of less than 1 $\mu$m ps$^{-1}$ \cite{Rozenman2018Long-RangeMicroscopy}. Recent work has also shown unexpected dependence of polariton wave packet width on  the microcavity quality factor \cite{Pandya2022TuningDelocalization} as well as the coexistence of diffusive and ballistic transport regimes controllable via the photonic content of the dominant wave packet component \cite{Balasubrahmaniyam2023FromExcitations}. Another key feature reported in recent work \cite{Xu2022UltrafastInteractions, Balasubrahmaniyam2023FromExcitations} is the renormalization of the ballistic transport velocity and diffusion constants of exciton-polaritons induced by dynamical and static disorder.

Inspired by promising experimental findings, several theoretical studies have emerged aiming to clarify and optimize the mechanism underlying polariton-assisted transport \cite{Feist2015ExtraordinaryCoupling, Schachenmayer2015Cavity-EnhancedExcitons,Gonzalez-Ballestero2016UncoupledProperties, Du2018TheoryTransfer, Reitz2018EnergyConfigurations, Schafer2019ModificationChemistry, Scholes2020,li2021collective, Gettapola2021DirectionalNanostructures, Engelhardt2022PolaritionMicrocavities, ribeiro2023vibrational}.
Models based on a one-dimensional molecular chain coupled to a single cavity mode showed that the collective coupling between molecules and the photonic material might overcome generic disorder-induced transport suppression \cite{Feist2015ExtraordinaryCoupling, Schachenmayer2015Cavity-EnhancedExcitons}. It has also been proposed that hopping and cavity-mediated energy transfer formed independent transport channels \cite{Feist2015ExtraordinaryCoupling, Allard2022Disorder-enhancedPhotons}. The role of dark states, i.e., states with small or zero photonic content, has also been examined. It was proposed that these states, unlike noninteracting molecular states outside a microcavity, can be spatially delocalized and therefore contribute to efficient energy transport \cite{Gonzalez-Ballestero2016UncoupledProperties, Botzung2020DarkCavity}. In spite of these compelling insights, most theoretical work so far has been done using single photon mode theories, where molecules inside the cavity are coupled to a single spatially homogeneous photon mode. Earlier work by Agranovich and Gartstein investigated polariton propagation along a one-dimensional multimode cavity establishing that disorder tends to localize polariton modes with nearly zero wave vector \cite{agranovich2007nature}. 
The importance of a multimode description of the electromagnetic field has become clear in recent studies \cite{Tichauer2021Multi-scaleRelaxation, Allard2022Disorder-enhancedPhotons, Ribeiro2022MultimodeFluctuations, Engelhardt2022PolaritionMicrocavities, ribeiro2023vibrational}. For example, it has been shown that significantly different dynamics occur in the presence of a more realistic radiation field including multiple degrees of freedom  \cite{Tichauer2021Multi-scaleRelaxation, Ribeiro2022MultimodeFluctuations}. Nevertheless, a quantitative analysis is still lacking on how predictions of dynamical exciton phenomena depend on the various choices that need to be made in the computational investigation of multimode polaritonic materials (e.g., finite system size and number of cavity modes). In addition, phenomenological questions remain open on the timescales associated with exciton transport phenomena such as ballistic, diffusive, subdiffusive (transient and long-time localization), and disorder-enhanced transport in optical cavities.

In this work, we analyze the coherent propagation of exciton wave packets in a lossless polaritonic wire to study multimode dynamics and numerical accuracy. We compute the time evolution of exciton wave packets and examine how the results depend on the system size and number of cavity modes. Next, we discuss which photon modes provide the dominant contribution to the exciton transport and how this property is affected by the initial state preparation, light-matter interaction strength, and static energetic disorder. Although energetic disorder introduces nontrivial dynamical effects, we are able to present a simple energy-based criterion to determine the most relevant set of photon modes. Finally, we discuss the most prominent dynamical features observed in our simulations. Since our model does not include dissipative effects, our conclusions can be seen as limiting or upper-bound expectations. Nevertheless, we anticipate the analysis presented here provides new insight into dynamical aspects of exciton evolution under strong light-matter coupling and recommendations for future theoretical model development. 

\section{Theory and Computation}
\subsection{Polaritonic Wire}
\par The polaritonic wire model, illustrated in Figure 1, describes a perfectly reflective cuboid microcavity with confinement lengths $L_x$, $L_y$, and $L_z$, where $L_x$ is the long dimension ($L_x \gg L_z, L_y$). This ideal cavity confines the EM field along the $y$ and $z$ coordinates while we impose periodic boundary conditions along the $x$ direction. Molecules are fixed at sites distributed along $x$ with an average intermolecular distance $a$. In a system with $N_M$ molecules, the length along $x$ is $L_x = N_M a$, whereas $L_y$ and $L_z$ are fixed at 200 and 400 nm, respectively. 
The total Hamiltonian for this system can be separated into three components
\begin{align}
    \hat{H} = \hat{H}_\text{M} + \hat{H}_\text{L} + \hat{H}_\text{LM} \label{Hamiltonian}
\end{align}
representing matter ($\hat{H}_\text{M}$), light ($\hat{H}_\text{L}$), and their interaction ($\hat{H}_\text{LM}$).

\begin{figure}
    \centering
    \includegraphics[width=\textwidth]{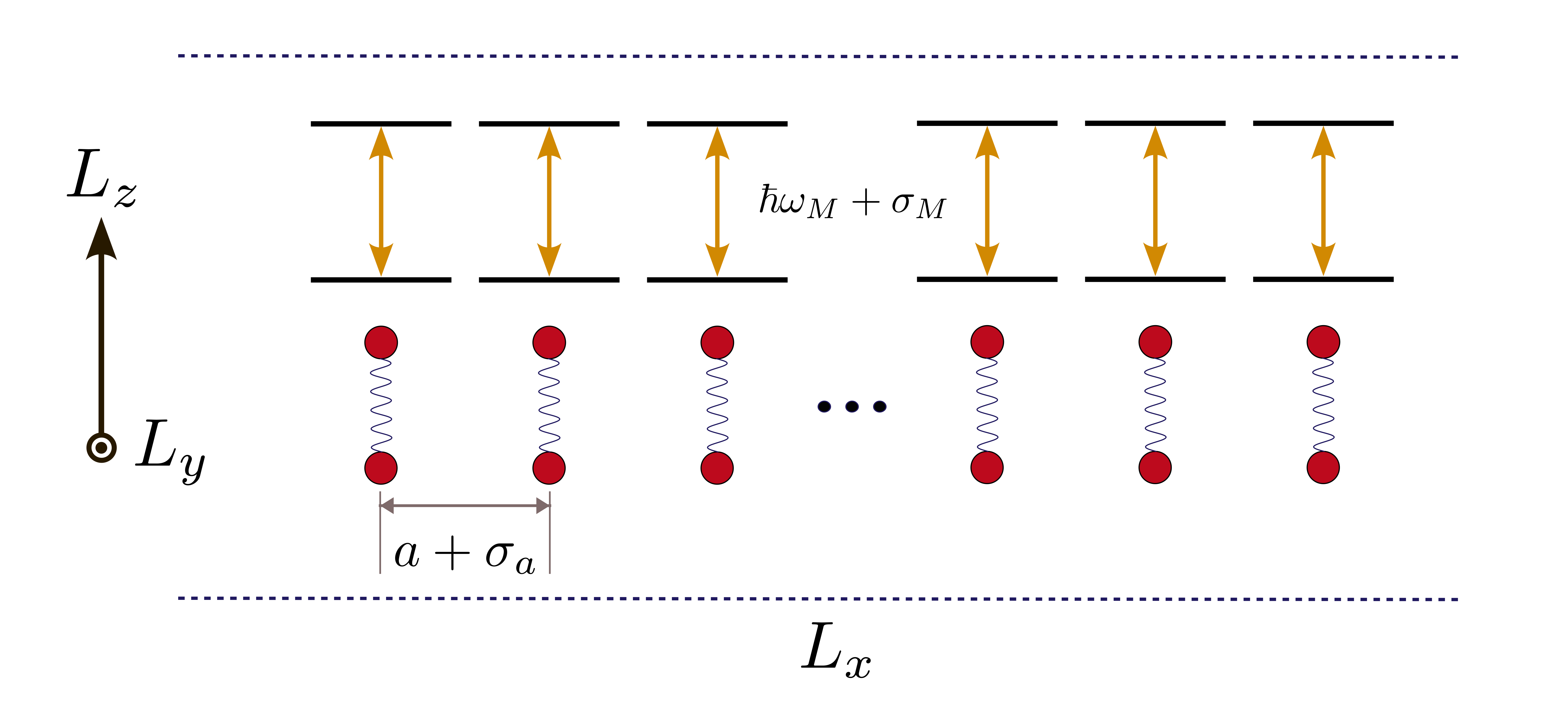}
    \caption{Model of a lossless cuboid microcavity (polaritonic wire) hosting $N_{\text{M}}$ two-level systems fixed at sites along $x$ representing molecules. The distance between sites ($a$) and the excitation energy ($\hbar\omega_M$) of each molecule includes disorder drawn from a normal distribution with corresponding standard deviations $\sigma_M$ and $\sigma_a$. Throughout this paper, $L_y = 200$ nm and $L_z = 400$ nm are employed.}
    \label{fig:wire}
\end{figure}

\subsection{Cavity Modes}

\par For the sake of simplicity, we assume the molecular transition dipoles are parallel to the transverse-electric radiation modes of the photonic wire. The cavity radiation field satisfies periodic boundary conditions along $x$ and Dirichlet conditions along $y$ and $z$. Consequently, the allowed wave vectors are characterized by three quantum numbers
\begin{align}
    \mathbf{k} = \left(\frac{2\pi m_x}{L_x},\; \frac{\pi n_y}{L_y},\; \frac{\pi n_z}{L_z} \right)\,,
\end{align}
where $m_x \in \mathbb{Z}$ and $n_y, n_z \in \mathbb{N}_{>0}$. Since the quantization length is small along $y$ and $z$, the energy gap between different values of $n_y$ and $n_z$ is large. Hence, we consider only the lowest energy band, i.e., $n_y = n_z = 1$. Defining the quantities
\begin{align}
    q &\equiv k_x = \frac{2\pi m_x}{L_x}\,, \label{qdef}\\
    q_0 &\equiv \sqrt{k_y^2 + k_z^2} = \sqrt{\left(\frac{\pi}{L_y}\right)^2 + \left(\frac{\pi}{L_z}\right)^2}\,, \label{q0}
\end{align}
we can fully describe each cavity mode by its value of $q$. The Hamiltonian of the EM field can be written as
\begin{align}
    \hat{H}_\text{L} = \sum_q \hbar \omega_q \hat{a}_q^\dagger \hat{a}_q\,, \label{HL}\\
    \omega_q = \frac{c}{\sqrt{\epsilon}}\sqrt{q_0^2 + q^2} \,,\label{cmodes}
\end{align}
where $\hbar$ is the reduced Planck constant, $c$ is the speed of light, $\epsilon$ is the relative static permittivity, and $\hat{a}^\dagger$ and $\hat{a}$ are bosonic creation and annihilation operators, respectively. The sum over cavity modes in Eq. \ref{HL} is, in principle, infinite. However, the importance of a particular mode becomes negligible as it grows highly off-resonant with the transitions of interest. Therefore, the number of cavity modes ($N_c$) is truncated by choosing a cutoff value $\hbar\omega_{q_\text{max}}$. In general, the $N_c$ modes include positive and negative $q$ values representing counterclockwise and clockwise waves, but we also consider the effect of retaining only one direction. The relative permittivity is chosen as $\epsilon = 3$, compatible with organic microcavities, and combined with the chosen wire geometry, yields a minimum cavity energy $\hbar\omega_{q=0}$ = 2.00 eV.

\subsection{Molecular Hamiltonian}

Each molecule along the wire is described as a two-level system with mean excitation energy $\hbar \omega_M$ and an average intermolecular distance of $a$. Static disorder is introduced in the system by allowing molecular excited-stated energies and positions to vary following the normal distributions
\begin{align}
    P(E_n) &= \frac{1}{\sigma_M\sqrt{2\pi}}\exp\left[{-\frac{1}{2}\left(\frac{E_n-\hbar\omega_M}{\sigma_M}\right)^2}\right]\,, \label{PE}\\
    P(x_n) &= \frac{1}{\sigma_a\sqrt{2\pi}}\exp\left[{-\frac{1}{2}\left(\frac{x_n-(n-1)a}{\sigma_a}\right)^2}\right]\,, \label{PX}
\end{align}
where $E_n$ and $x_n$ are the energy and position of the $n$-th molecule, respectively. The standard deviations $\sigma_M$ and $\sigma_a$ characterize the static disorder of the system. We assume intermolecular interactions are weak enough that any bare energy transfer occurs on a much longer timescale than probed by our simulations and, hence, can be ignored. The matter component of the Hamiltonian is written simply as 
\begin{align}
    \hat{H}_\text{M} = \sum_{n}^{N_M} E_n \hat{b}_n^\dagger \hat{b}_n\,,
\end{align}
with $E_n$ sampled from Eq. \ref{PE} and $\hat{b}_n^\dagger$ ($\hat{b}_n$) representing an operator that promotes (demotes) the $n$-th molecule to (from) its excited state (i.e., $\hat{b}_n = \ket{0}\bra{1_n}$). 

\subsection{Light-Matter Interaction}

\par Employing the Coulomb gauge in the rotating wave approximation and neglecting the diamagnetic contribution, the interaction between light and matter can be written in terms of the collective light-matter interaction strength (Rabi splitting, $\Omega_R$) as follows:
\begin{align}
    \hat{H}_\text{LM} = \sum_n^{N_M}\sum_q -i\frac{\Omega_R}{2}\sqrt{\frac{E_n}{N_M\hbar\omega_q}}\left(e^{iqx_n}\hat{b}_n^\dagger \hat{a}_q - e^{-iqx_n}\hat{a}_q^\dagger \hat{b}_n\right)\,. 
\end{align}
Since $\Omega_R = \mu\sqrt{\hbar\omega_0\rho/2\epsilon}$, choosing a value of $\Omega_R$ and density ($\rho = N_M/L_xL_yL_z$) implies a transition dipole moment ($\mu$). We assume that all molecular transition dipoles are aligned with the polarization of the cavity modes; therefore, different interactions between molecules and photons arise only from their relative energy and the varying electric field amplitude along the wire for $|q| > 0$. 

\subsection{States and time evolution}

\par The initial states of the simulations consist of purely molecular exciton Gaussian wave packets placed at the center of the wire ($N_Ma/2$). In the uncoupled basis, these wave packets can be represented as
\begin{align}
    \ket{\psi(0)} = \frac{1}{Z}\sum_n \exp\left[-\frac{(x_n-\frac{1}{2}N_Ma)^2}{2\sigma_x^2} + i\bar{q}_0x_n\right] \ket{1_n}\otimes\ket{0}\,, \label{gaussianwvp}
\end{align}
where $Z$ is a normalization constant, $\sigma_x$ is the initial spread of the wave packet (Eq. \ref{x2}), and $\ket{1_n}\otimes\ket{0}$ represents a state where the $n$-th molecule is in its excited state while all other molecules and cavity modes are in the ground state. The mean initial exciton momentum is given by $\bar{q}_0$ and, unless otherwise noted, it is taken as zero in all simulations. The wave packet dynamics is obtained by first diagonalizing the Hamiltonian matrix  (Eq. \ref{Hamiltonian}), and constructing the time-evolved wave packet exactly as $\ket{\psi(t)} = e^{-i\hat{H} t / \hbar } \ket{\psi(0)}$.

We investigate the dynamics within the one-excitation manifold. Therefore, our results are mostly relevant in the dilute limit (weak pumping) scenario where the exciton density is small, and nonlinearities can be ignored. Note that in disordered microcavities, nearly pure molecular initial states can be constructed via resonant excitation of weakly coupled modes.  

To obtain a measure of the number of molecules over which the wave packet extends, we compute the wave packet width $d(t)$ defined as the root mean square displacement divided by the average intermolecular distance
\begin{align}
    d(t) &= \frac{\sqrt{\langle x(t)^2 \rangle}}{a}\,, \label{d}\\
    \langle x^2(t) \rangle &= \frac{1}{P_\text{mol}(t)}\sum_n^{N_M} |c_n(t)|^2(x_n - x_0)^2\,, \label{x2}\\
    P_\text{mol}(t) &= \sum_n^{N_M} |\bra{1_n}\ket{\psi(t)}|^2 = \sum_n^{N_M} |c_n(t)|^2\,,
\end{align}
where $c_n(t) = \langle 1_n | \psi(t) \rangle$ are the local molecular amplitudes and $P_\text{mol}(t)$ is included as normalization factor in Eq. \ref{x2} so the molecular wave packet width is computed from the conditional probability of finding the exciton on a given site.

All computations were performed using our prototype package \textsc{PolaritonicSystems.jl} \cite{psys}. Double-precision complex numbers were used for representing state vectors. Matrix operations were carried out with Intel MKL and LAPLACK backend. Random numbers sampled from normal distributions were generated using the \textsc{Distributions.jl} package\cite{Besancon2019}. Figures were produced using the \textsc{Makie.jl} plotting ecosystem \cite{makie}.

\section{Results}
\subsection{Size effects}

\par
\begin{figure}
    \centering
    \includegraphics[width=\textwidth]{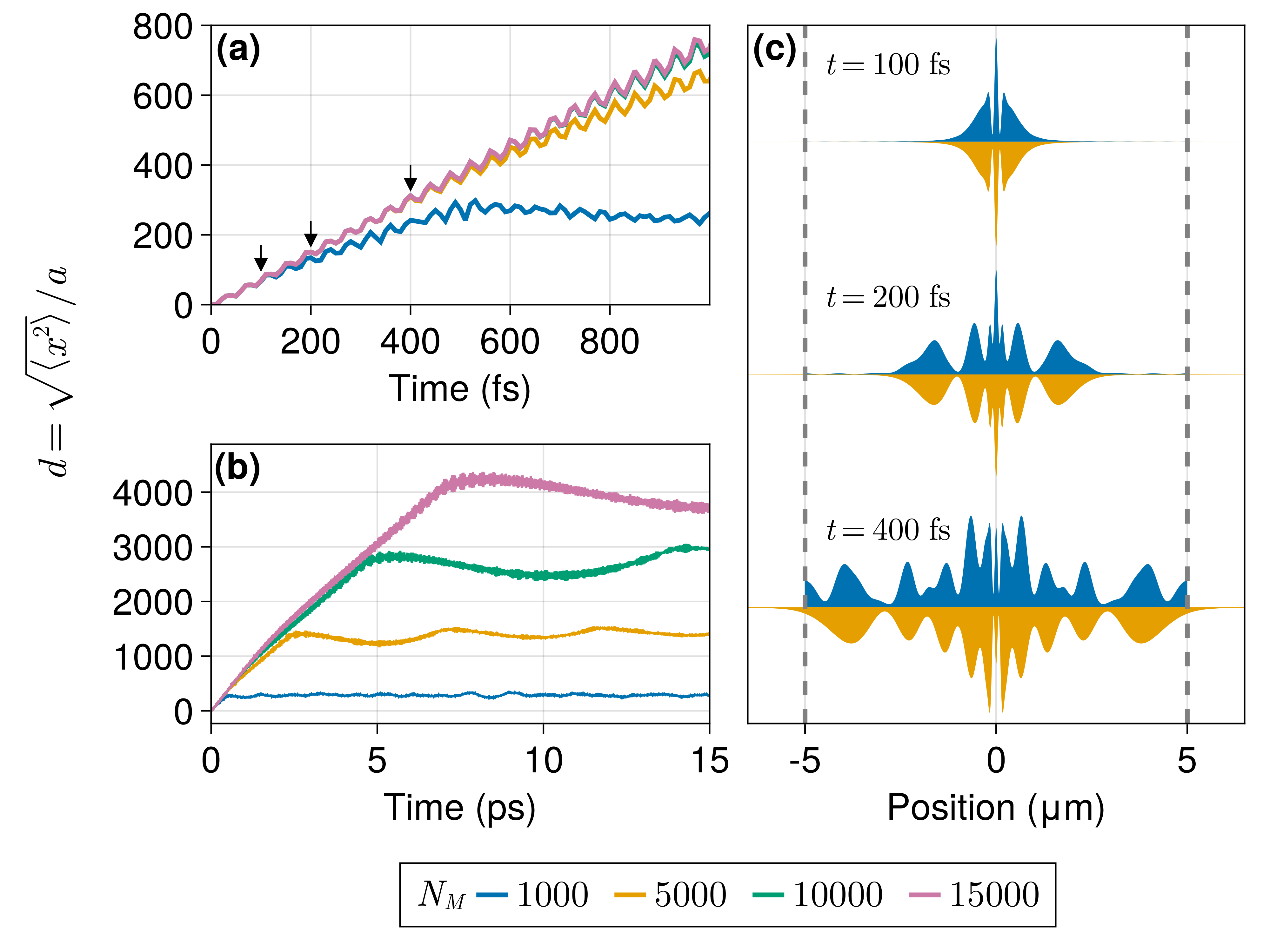}
    \caption{Wave packet width ($d$) over short (a) and long (b) propagation times for several system sizes with no disorder. (c) exciton wave packet shapes (arbitrary scale) for $N_M = 1000$ and $5000$ at selected time steps, indicated by arrows on (a). Dashed lines indicate the same point in the $N_M = 1000$ (circular) wire. The radiation field was modeled using 1601 cavity modes, and the Rabi splitting was set to 0.1 eV. Molecules are positioned 10 nm apart from each other. The lowest cavity mode is in resonance with the molecular excitation energy of 2.0 eV. The initial wave packet was prepared with $\sigma_x = 60$ nm ($d(0) = 6$ molecules). Cavity cutoff energies are 57.30, 11.63, 6.07, 4.31, and 3.49 eV, in increasing order of $N_M$.}
    \label{fig:ord_prop}
\end{figure}

 In most experiments, the collective strong coupling regime is achieved using a macroscopic number of molecules in optical microcavities \cite{agranovich2003cavity, Ebbesen2016HybridPerspective}. For the sake of computational feasibility, simulations are performed on a much smaller system size. To assess how this reduction affects exciton dynamics, we simulated the propagation of wave packets at various system sizes in the absence of disorder. The results presented in Fig. \ref{fig:ord_prop} refer to simulations with variable $L_x = N_M a$, fixed photon number $N_c = 1601$ corresponding to variable cutoff energies for the cavity, Rabi splitting $0.1$ eV, and initial wave packet width $\sigma_x = 60$ nm. The large number of photon modes employed here was chosen to ensure convergence, the sensitivity of our results with respect to this parameter is explored in the following subsections. The evolution of the wave packet width $d(t)$ for multiple system sizes is shown for short and long propagation times in Figs. \ref{fig:ord_prop}(a) and (b), respectively, while snapshots of wave packets at selected times are shown in Fig. \ref{fig:ord_prop}(c). 

In every examined case, we see a linear ballistic-like transport with small-amplitude oscillations around the main profile. These oscillations have a period compatible with the corresponding Rabi splitting, indicating that they arise due to the periodic exchange of energy between light and matter. At longer times, the exciton width reaches a plateau around which it oscillates. This apparent localization is a finite system size effect, as in the absence of disorder, the wave packet motion in an infinite system is unbounded. At propagation times under 200 fs, the same time-linear profile is found for all selected system sizes. The worst case is seen for $N_M = 1000$  where the exciton slows down significantly as the wave packet extends over all molecules within approximately 400 fs (as shown in Fig. \ref{fig:ord_prop}(c), see also SI Fig. S1). Note that different system sizes will imply different photon density of states, which may be a source of error. Nevertheless, our computed values of $d$ and wave packet shapes are well in agreement for all different sizes up to localization due to finite size. Since disorder induces wave packet localization \cite{anderson1958absence,Allen1998EvolutionChain}, we expect it will lessen the finite-size effects responsible for the deviations observed in Fig. 2. Therefore, these results suggest that transport properties can be probed with a small number of molecules as long as the system is longer than the exciton localization length scale or the probing time is earlier than the localization time scale.

\subsection{Electromagnetic field truncation}

\par Although recent works have highlighted the importance of a multimode description of the cavity radiation field, \cite{Tichauer2021Multi-scaleRelaxation,Ribeiro2022MultimodeFluctuations,Allard2022Disorder-enhancedPhotons} as far as we are aware, a quantitative analysis on how the accuracy of molecular observables depends on the number of cavity modes has not yet been presented. To assess the convergence of the exciton dynamics with respect to the number of photon modes ($N_c$), we compute the time-dependent exciton width (Eq. \ref{d}) using different cavity cutoffs $q_\text{max}$ (Eq. \ref{cmodes}). For any choice of cutoff, all modes with $q$ satisfying $|q| < q_\text{max}$ are retained. A subsequent section examines the consequences of neglecting $q<0$ modes. 

\begin{figure}
    \centering
    \includegraphics[width=\textwidth]{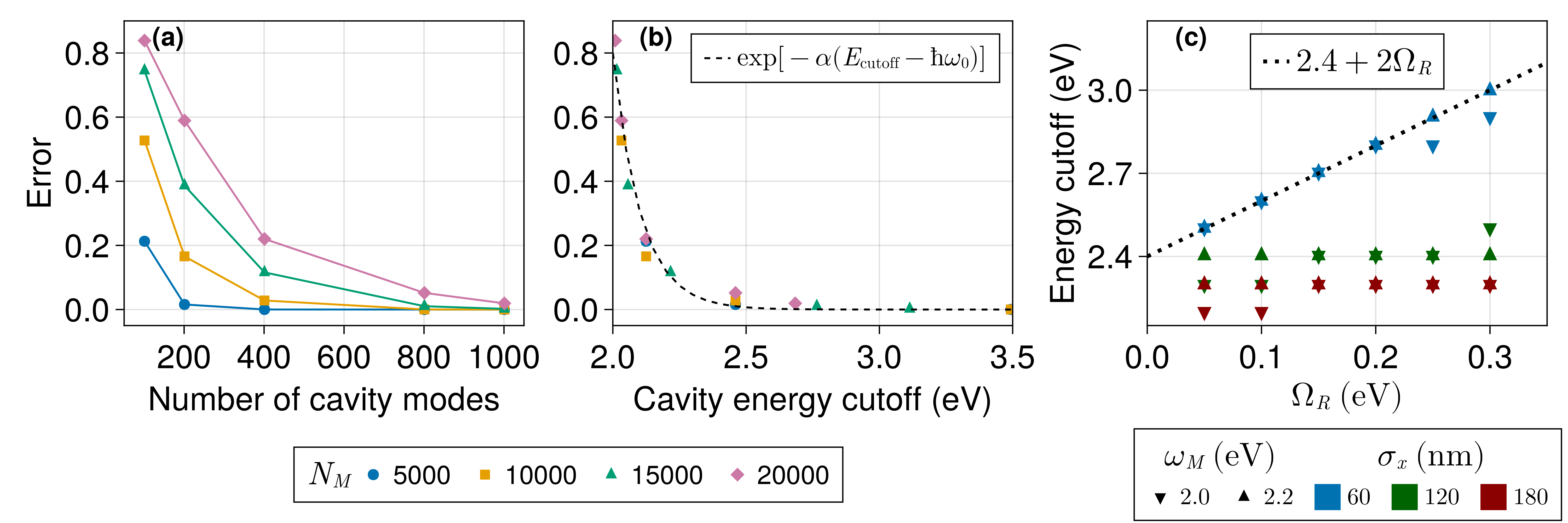}
    \caption{Error in the wave packet width (Eq. \ref{error}) without disorder for several system sizes as a function of the number of cavity modes (a) and energy cutoff value (b). The error was computed over an interval of 5 ps with time steps of 10 fs. Rabi splitting ($\Omega_R$) was set to 0.1 eV. Molecules are positioned 10 nm apart from each other. The lowest cavity mode is in resonance with the molecular excitation energy ($\omega_M$) of 2.0 eV. The initial wave packet was prepared with $\sigma_x = 60$ nm ($d(0) = 6$ molecules). (c) Estimated energy cutoff necessary for a truncation error under 0.01 as a function of $\Omega_R$  for several values of $\sigma_x$ and $\omega_M$.}
    \label{fig:error_nodis}
\end{figure}

We employ the following global (integral) measure of wave packet propagation error due to cavity mode truncation 
\begin{align}
    \text{error}(N_c) = \frac{1}{t}\int_0^{t} \frac{|d(t) - d_\text{ref}(t)|}{d_\text{ref}(t)} dt \approx \frac{1}{\mathcal{N}} \sum_i^\mathcal{N} \frac{|d(t_i) - d_\text{ref}(t_i)|}{d_\text{ref}(t_i)}. \label{error}
\end{align}
where $d_\text{ref}$ is taken from a computation using a large number of photon modes ($N_c = 1601$) sufficient to achieve converged dynamics. In Fig. \ref{fig:error_nodis}(a), we show that, without disorder, in the best case scenario with $N_M = 5000$, a large $N_c > 200$ is necessary to reduce the numerical error measured by Eq. \ref{error} substantially. An increase in the size of the system leads to a larger error at fixed $N_c$ and slower convergence towards the exact result as $N_c$ increases. This happens because the photon energy spacing  $\Delta E_q  = E_{q_{n+1}}-E_{q_n}$ decreases with the system length (Eqs. \ref{qdef}-\ref{cmodes}). For instance, when $N_M = 5000$ the lowest 201 modes span 0.46 eV, while simulations with $N_M = 10^4$ and $N_M = 2 \times 10^4$ require 401 and 801 modes, respectively, to span the same energy range. When the truncation errors are presented as a function of the photon energy cutoff [Fig. \ref{fig:error_nodis}(b)], a general trend emerges, indicating exponential decay of the error with cavity cutoff. Hence, the energy range spanned by the included photon modes is the most relevant parameter of the truncation. Fig. \ref{fig:error_nodis}(c) shows the necessary upper energy cutoff to keep the truncation error below 0.01 as a function of $\Omega_R$. We found that detuning (redshifting the cavity) had little effect on this trend. However, the initial state spread affected the result significantly. In the worst case scenario, for a narrow initial wave packet ($\sigma_x = 60$ nm), we found that the required photon energy cutoff in the absence of disorder scales linearly as $2\Omega_R$. This factor can serve as a heuristic relationship to estimate the range of important photon modes. 

\par Disorder induces wave function localization \cite{anderson1958absence,agranovich2003cavity,agranovich2007nature} and potentially reduces the total photon content of the wave packet, e.g., for an exciton with $\sigma_x = 120$ nm and $\Omega_R = 0.1$ eV, the average photon content drops from 40\% to around 20\% when the relative energetic disorder is increased from $0.05\Omega_R$ to $0.2\Omega_R$ (see SI Fig. S19). At the same time, the exciton propagation in a disordered landscape is irregular, involving many scattering events that might require a more flexible description of the field that includes many degrees of freedom. In light of these considerations, and that disorder is an unavoidable feature of polaritonic materials, it is important to determine how the introduction of static disorder changes the accuracy of simulations performed with a finite number of cavity modes. 

\par Our model includes energetic and positional disorder as described in Eqs. \ref{PE} and \ref{PX}. As reported in earlier work \cite{litinskaya2006loss, litinskaya2008propagation, Ribeiro2022MultimodeFluctuations}, even small energetic disorder dominates over translational. Therefore, we fix the molecular position standard deviation to be $\sigma_a = 1$ nm in all calculations that include disorder. To assess the effects of field truncation, we set $\Omega_R = 0.1$ eV, $\sigma_x = 60$ nm, and $a = 10 ~\text{nm}$ as representative examples along with $N_M = 5000$ following our previous discussion of finite size effects. The truncation error is computed from Eq. \ref{error} using the average value of $d(t)$ obtained from 100 realizations for each probed $N_c$ and $\sigma_M$. The computed errors' uncertainty was obtained using linear propagation theory implemented in the Measurements.jl package\cite{Measurements.jl-2016}.

\par
\begin{figure}
    \centering
    \includegraphics[width=\textwidth]{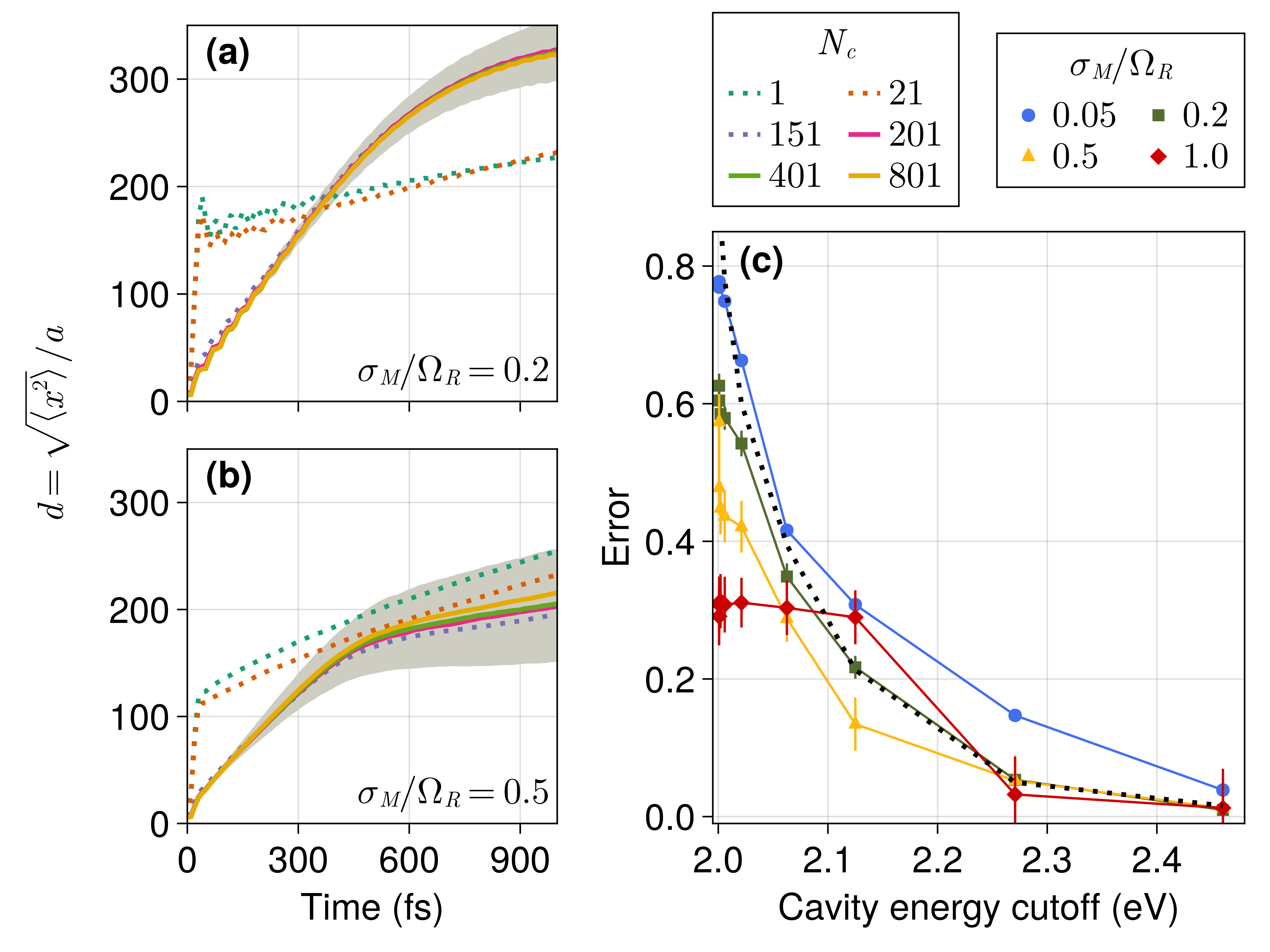}
    \caption{Wave packet width, $d(t)$, over time using different numbers of cavity modes ($N_c$) with relative disorder strength $\sigma_M/\Omega_R = 0.2$ (a) and $\sigma_M/\Omega_R = 0.5$ (b). Energy cutoffs are 2.00, 2.01, 2.27, 2.46, 3.49, and 6.07 eV, in increasing order of $N_c$. Dotted lines highlight trajectories with error above 0.05. (c)  
    Error in the wave packet width (Eq. \ref{error}) for several energetic disorder values and $\Omega_R = 0.1$ eV. The dotted black line shows the convergence observed without disorder. The lowest cavity mode is in resonance with the average molecular excitation energy of 2.0 eV. The wire contains 5000 molecules with $a = 10$ and $\sigma_a = 1$ nm. The initial wave packet was prepared with $\sigma_x = 60$ nm ($d(0) = 6$ molecules). The simulation was run for 1 ps with time steps of 10 fs. Values shown are averages over 100 realizations, and error bars are propagated from twice the standard deviation of $d$.}
    \label{fig:error}
\end{figure}

 The average exciton $d(t)$ profiles obtained with several $N_c$ values when the energetic disorder $\sigma_M$ is equal to 20\% and 50\% of $\Omega_R$ are presented in Fig. \ref{fig:error}(a) and \ref{fig:error}(b), respectively. The shaded region covers twice the standard deviation from $d_\text{ref}$ in both cases. The error analysis is more complex in the presence of disorder due to the stochastic nature of the system. Nonetheless, we can clearly distinguish mean trajectories that are qualitatively different from the reference. For instance, average trajectories with $N_c = 1$ and $N_c = 21$ exhibit significant deviation from the reference, indicating the qualitative incorrectness of the corresponding incomplete mode expansion. The mean $d(t)$ obtained with $N_c = 151$ lies just above our error threshold of 0.05 and, for the most part, is contained within two standard deviations of $d_{\text{ref}}(t)$. Errors as a function of cavity energy cutoffs for several disorder strengths are shown in Fig. \ref{fig:error}(c). In spite of some complicated features at inadequately low cutoff energies, the overall trend is not too different from that observed in the absence of disorder (Fig. \ref{fig:ord_prop}). Therefore, we believe the results obtained in the absence of disorder can be used to reliably estimate the number of cavity modes needed for a particular system.

\subsection{Photon weight distribution}

\par The finding that cavity modes within a broad energy range contribute to exciton propagation is at odds with the intuition that resonant processes must dominate the dynamics over sufficiently long times. To quantify the contribution of the various field modes to the exciton evolution, we analyzed the composition of the photonic part of the wave packet via the time-averaged relative mode weight distribution
\begin{align}
    W(q) = \frac{1}{W_\text{max}}\sum_t |\bra{q}\ket{\psi(t)}|^2, \label{mode_weight_eq}
\end{align}
where $\ket{q} =  \hat{a}_q^\dagger \ket{0}$ represents a state with one photon with wave vector $q$ and no molecular excitations, and $W_\text{max}$ is equal to  $\text{max}_q~\sum_t |\bra{q}\ket{\psi(t)}|^2$. The discrete sum over times ($t$) was performed numerically using time increments of 5 fs for a total period of 5 ps. 

\par
\begin{figure}
    \centering
    \includegraphics[width=\textwidth]{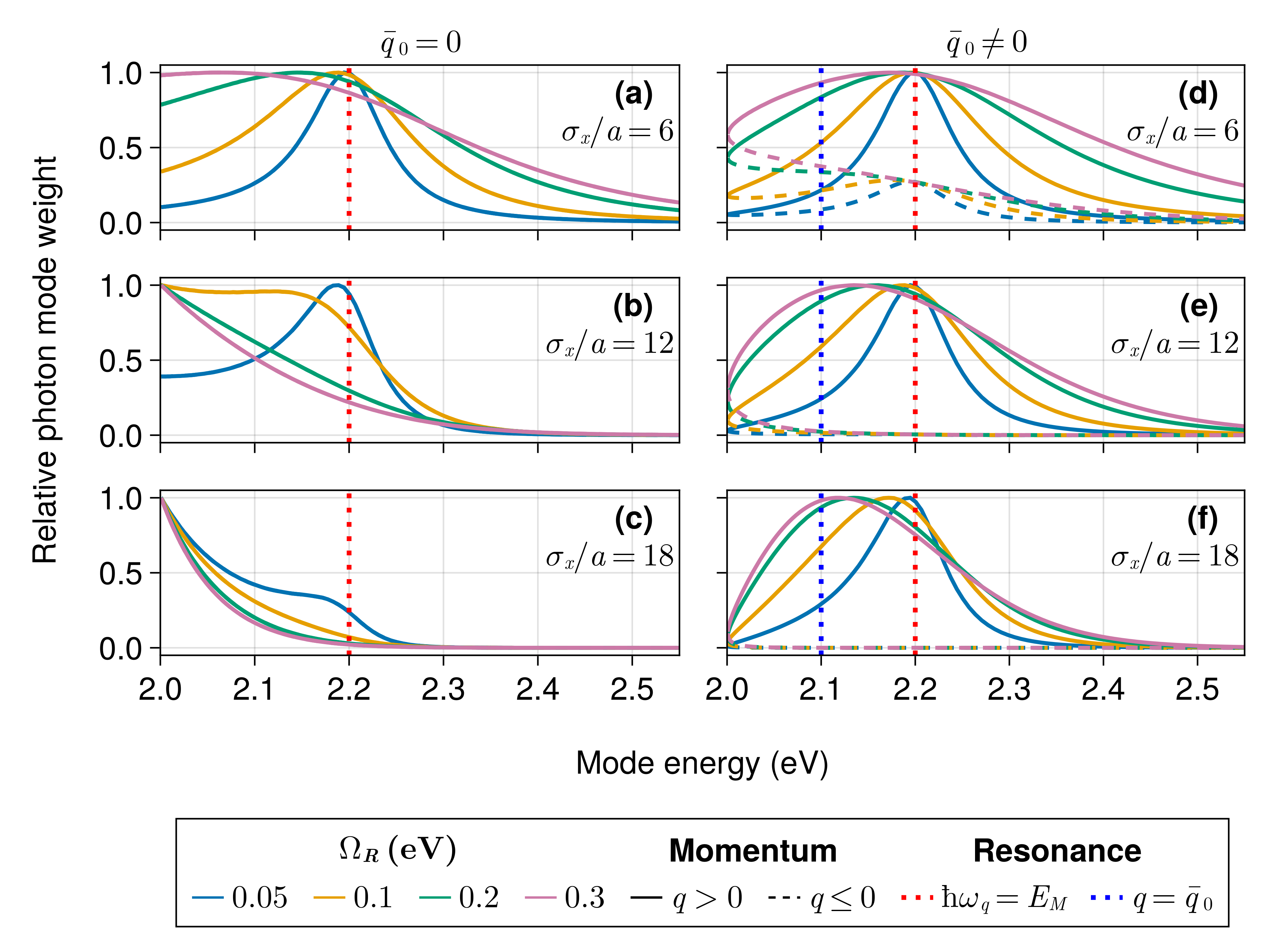}
    \caption{Cavity mode contribution measured by Equation \ref{mode_weight_eq} under no disorder. The computation was performed over 5 ps using a 5 fs time step. The number of molecules and cavity modes is set to 5000 and 401, respectively. The distance between molecules ($a$) is fixed at 10 nm, and the molecular excitation energy is 2.2 eV, indicated by the dotted red line. The mean exciton momentum ($\bar{q}_0$) is set to zero (a,b,c) or 0.00565 nm$^{-1}$ (d,e,f), which matches the momentum of the photon at 2.1 eV (dotted blue line). Modes with $q>0$ and $q\leq 0$ are represented by solid and dashed lines, respectively.}
    \label{fig:mw_ideal}
\end{figure}

 In Fig. \ref{fig:mw_ideal}, we present $W(q)$ obtained in the absence of disorder for various $\Omega_R$ and variable initial exciton widths $d(0)$ with $N_m = 5000$ and $N_c = 401$. In this subsection, we work with variable $\sigma_x$ and $\sigma_x$ because these are experimentally tunable quantities that significantly affect the time-averaged relative photon weight distribution. The mean molecular excitation energy was set to 2.2 eV (resonant with the photon modes $q_r$ such that $\hbar\omega_{q_r} = E_M$). We probed the time-averaged photon weight distributions for excitons with both vanishing and nonvanishing wave vectors ($\bar{q}_0 = 0$ and $\bar{q}_0 \neq 0$, respectively). 

\par The distinct choices of cavity-matter detuning ($E_0 - E_M = -0.2~\text{eV}$) and mean exciton wave vector examined in this section relative to the prior will allow us to reveal an interesting interplay between the competing biases of the photon weight distribution towards distinct wave vectors $q$ satisfying $q = q_r$ (energy resonance) and $q = \bar{q}_0$ (quasimomentum matching). The competition between $q_r$ and $\bar{q}_0$ can be  demonstrated analytically in the infinite system limit where $L_x \rightarrow \infty$ and simultaneously $N_M = N_c \rightarrow \infty$. In this case, for a system without disorder, the time-average photon mode probability distribution generated by an initial exciton Gaussian wave packet in a polaritonic wire can be approximated by (see SI for derivation):
\begin{align}
    \lim_{t \rightarrow \infty} W(q) \propto  \Pi_{qL} (1-\Pi_{qL}) e^{-\sigma_x^2(q-\bar{q}_0)^2}, \label{eq:pcq}
\end{align}
where $\Pi_{qL}$ total molecular contents of the lower polariton mode with wave vector $q$. Note that $W(q)$ is proportional to a product of two competing terms. The exponential factor favors the photon modes with the same mean quasimomentum $\bar{q}_0$ as the initial wave packet with typical fluctuations of size $1/\sigma_x$. However, the bias towards the initial momentum might be irrelevant if the prefactor $\Pi_{qL} (1-\Pi_{qL})$ is very small at $q = \bar{q}_0$. In fact, this prefactor is maximized when $\Pi_{qL} = \dfrac{1}{2}$, which happens at resonance $q = q_r$. It follows that, when $\Omega_R \ll |\hbar\omega_{\bar{q}_0} - E_M|$ and $\sigma_x < 1/{|q_r-\bar{q}_0|}$, the prefactor $\Pi_{qL} (1-\Pi_{qL})$ approaches zero near $q = \bar{q}_0$; therefore, $W(q)$ will be maximized around $q = q_r$. In the opposite limit, where $\Omega_R \gg |\hbar \omega_{\bar{q}_0}-E_M|$, $\Pi_{qL} (1-\Pi_{qL})$ is appreciable and varies slowly around $q = \bar{q}_0$, causing $W(q)$ to be maximized at $q = \bar{q}_0$. 

In summary, the energy resonance condition ($q = q_r$) will dominate when the Rabi splitting is not too large and the exciton is compact. Conversely, if the Rabi splitting is large enough, quasimomentum matching ($q = \bar{q}_0$) is expected to determine the most important photon modes. 

The photon weights shown in Fig. \ref{fig:mw_ideal} reflect the observations provided above based on Eq. \ref{eq:pcq} while also revealing a wide distribution of off-resonant modes (relative to the mean molecular excitation energy) contributing significantly to the overall dynamics. For instance, increasing $\Omega_R$ leads to a broader photon weight distribution with the corresponding peaks shifting towards the modes with $q = \bar{q}_0$ (2.0 eV when $\bar{q}_0 = 0$ and 2.1 eV when $\bar{q}_0 \neq 0$). Notably, in every case, for sufficiently large $\Omega_R$ the maximum of the photon weight distribution matches the mean exciton wave vector, while the cavity modes in resonance with the molecular system play a lesser role in the dynamics. Correspondingly, we see that when the exciton possesses a nonvanishing initial momentum $\bar{q}_0 > 0$, the contribution of photons with $q < 0$ is strongly suppressed. Further, the width of the photon weight distribution in $q$ space decreases when the initial state becomes more delocalized, as seen comparing Fig. \ref{fig:mw_ideal}(d) to (f). This feature is in accordance with Eq. \ref{eq:pcq} and stems from the uncertainty principle.

\par In the presence of disorder, Eq. \ref{eq:pcq} does not hold, and the total (photon + exciton) wave packet quasimomentum is not conserved anymore. The non-singular distribution of molecular transition energies implies that a range of photon modes will be nearly resonant with the molecular system, leading to a broader and more complex photon weight distribution. If the disorder is small enough, e.g., $\sigma_M = 0.005 eV$, the $W(q)$ profiles are in good agreement with the previous discussion [see SI Fig. 24(a)-(c)]. 

\par
\begin{figure}
    \centering
    \includegraphics[width=\textwidth]{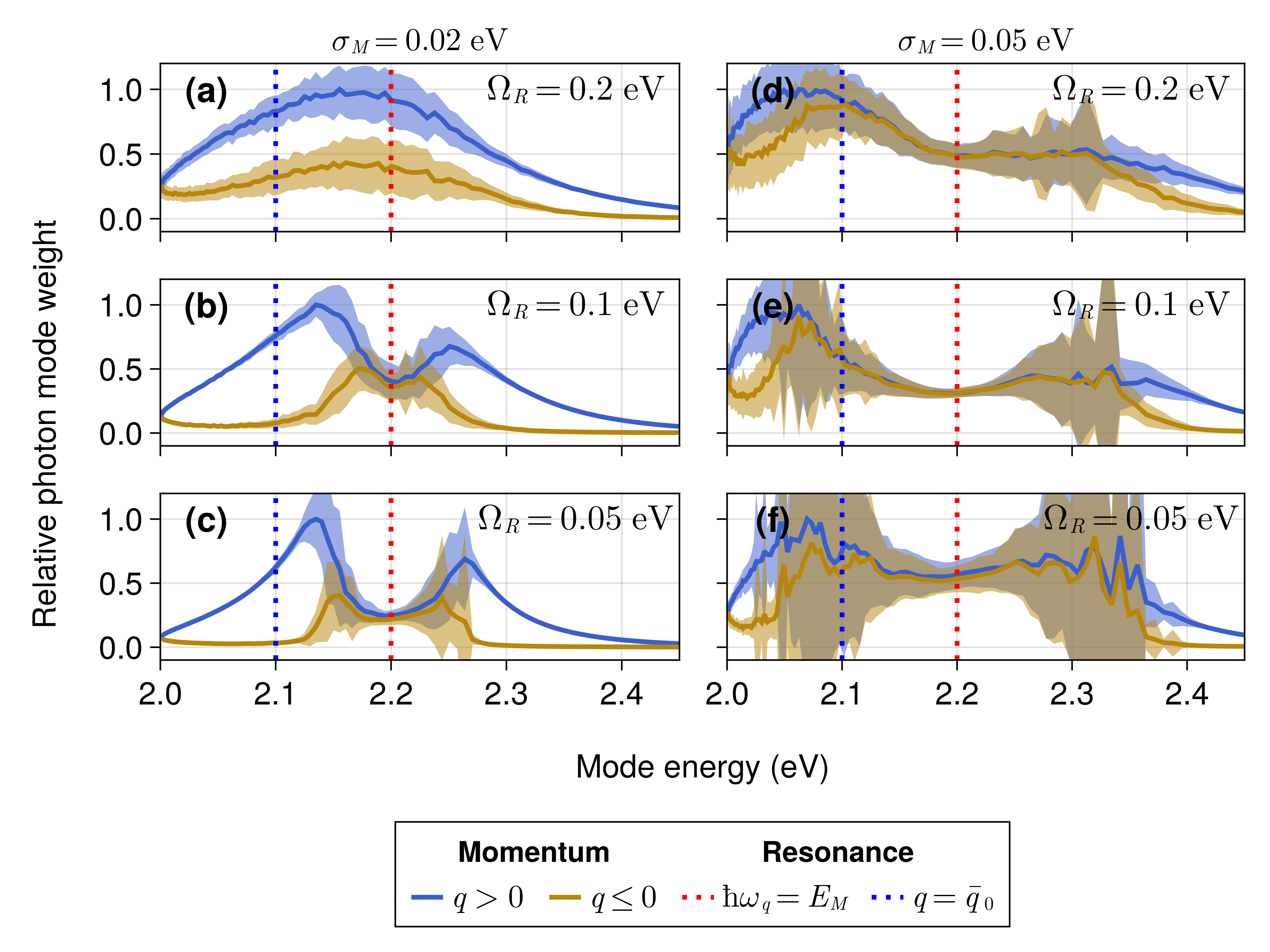}
    \caption{Cavity mode contribution measured by Equation \ref{mode_weight_eq} under energetic disorder of $\sigma_M = 0.02$ eV (a, b, c) and $\sigma_M = 0.05$ eV (d, e, f).  The wire was modeled with 401 cavity modes and 5000 molecules with intermolecular separations drawn from Eq. \ref{PX} using $a = 10$ and $\sigma_a = 1$ nm.  The average molecular excitation energy is 2.2 eV (dotted red line). The initial wave packet was prepared with $\sigma_x = 120$ nm ($d(0) \approx 12$ molecules) and an effective exciton momentum $\bar{q}_0 \approx$ 0.00565 nm$^{-1}$, which matches the momentum of the photon at 2.1 eV (dotted blue line). Modes with $q>0$ and $q\leq 0$ are represented in blue and yellow, respectively. The computation was performed over five ps using a five fs time step. Band plots cover one standard deviation around the average values of 100 realizations.}
    \label{fig:mw_dis}
\end{figure}

 Photon weight distributions under stronger disorder are shown in Fig. \ref{fig:mw_dis} for a wave packet with an initial width $\sigma_x/a = 12$ and $\bar{q}_0 > 0$. At the smallest relative disorder strength, $\sigma_M/\Omega_R = 0.1$ [Fig. \ref{fig:mw_dis}(a)], we observe similar features with only slight changes relative to those of Fig. \ref{fig:mw_ideal} (d)-(f), specifically the competition between the resonant quasimomentum $q_r$ and the initial exciton wave vector $\bar{q}_0$ leading to an average photon weight maximum at $ \bar{q}_0 <q < q_r$ [Fig. \ref{fig:mw_dis}(a)]. Similarly, the photon modes with $q < 0$ are suppressed relative to $q > 0$ even when $\sigma_M = 0.02$ and $\Omega_R = 0.05$ [Fig. \ref{fig:mw_dis} (c)]. However, when the energetic disorder is increased to 0.05 eV, we find nearly equal weights for positive and negative photon wave vectors at all examined values of $\Omega_R$ [Fig. \ref{fig:mw_dis}(d)-(f)]. This can be explained by the strong scattering induced by the disorder potential on the wave packet, which randomizes its $W(q)$ distribution. Moreover, the enhancement of coherent backscattering, which is a generic feature of the propagation of waves in disordered media, \cite{akkermans2007mesoscopic} favors a $W(q)$ distribution that is symmetric with respect to the inversion of $q$.

\par The most remarkable feature observed in Fig. \ref{fig:mw_dis} is the relative suppression of the photon modes nearly resonant with the center of the molecular excited-state energy distribution. Specifically, a dip at $q = q_r$ ($\hbar\omega_{q} = 2.2~\text{eV}$) can be clearly observed when $\sigma_M/\Omega_R$ is greater than $10\%$ [Figs. \ref{fig:mw_dis}(b), (c), (d), and (e)] and begins to fade at stronger disorder  $\sigma_M/\Omega_R = 1$ [Fig. \ref{fig:mw_dis} (f)]. In general, the time-average photon weight can be expressed as $W(q) \propto \sum_\chi |\langle q | \chi \rangle \langle \chi | \psi (0) \rangle|^2$, where the sum is over all eigenstates $\chi$ of the light-matter Hamiltonian (see SI). This means that a photon mode will have a large weight if it can overlap with eigenstates ($\chi$) that contribute significantly to the initial state ($\psi(0)$). A possible explanation for the increase in $W(q < q_r)$, especially with $\sigma_M = 0.05$ eV, may come from the fact that low $|q|$ photon modes provide a greater contribution to localized polaritons\cite{agranovich2003cavity, agranovich2007nature, litinskaya2006loss, litinskaya2008propagation}. In turn, these localized polaritons might have greater overlap with the exciton initial state than the eigenstates with significant $|q|\gg 0$ photon content. Future work will address this and other hypotheses for the $W(q > q_r)$ case. 

\par Another interesting feature observed in Fig. \ref{fig:mw_dis}(d)-(f) is the greater standard deviation observed for $q > q_r$ and $q < q_r$ relative to $q_r$. This is likely a byproduct of the greater fluctuations in both the number of  molecules with $E_n \neq \hbar\omega_{q_r}$ and the light-matter matrix elements  $(E_n/\hbar\omega_q)^{1/2}$ (for the interaction between molecule $n$ and cavity mode $q$) which arises in simulations with static energetic disorder. 

\par Overall, disorder significantly decreases the dominance of photon modes with a particular wave vector or energy, leading to a flatter photon weight profile relative to the absence of disorder. This emphasizes the importance of utilizing a flexible and unbiased description of radiation, as disorder makes the choice of photon modes less obvious. Nevertheless, in agreement with the non-disordered results discussed earlier, highly off-resonant modes ($\hbar\omega_q > 2.4 + 2\Omega_R$ eV) were observed to have negligible weight on the dynamics, further validating our energy cutoff criterion.

\subsection{Exciton wave packet dynamics in unidirectional polaritonic wires}

\par In the previous section, we investigated the convergence of our results with respect to the number of cavity modes in order the find the optimal set of photonic modes required to model exciton wave packet dynamics in photonic wires accurately. Another possible strategy to minimize the complexity of our model is to remove the double degeneracy of the cavity modes and retain only photons with $q \geq 0$. In this brief section, we examine the practical consequences of this alternative mode truncation approach.

\par
\begin{figure}
    \centering
    \includegraphics[width=\textwidth]{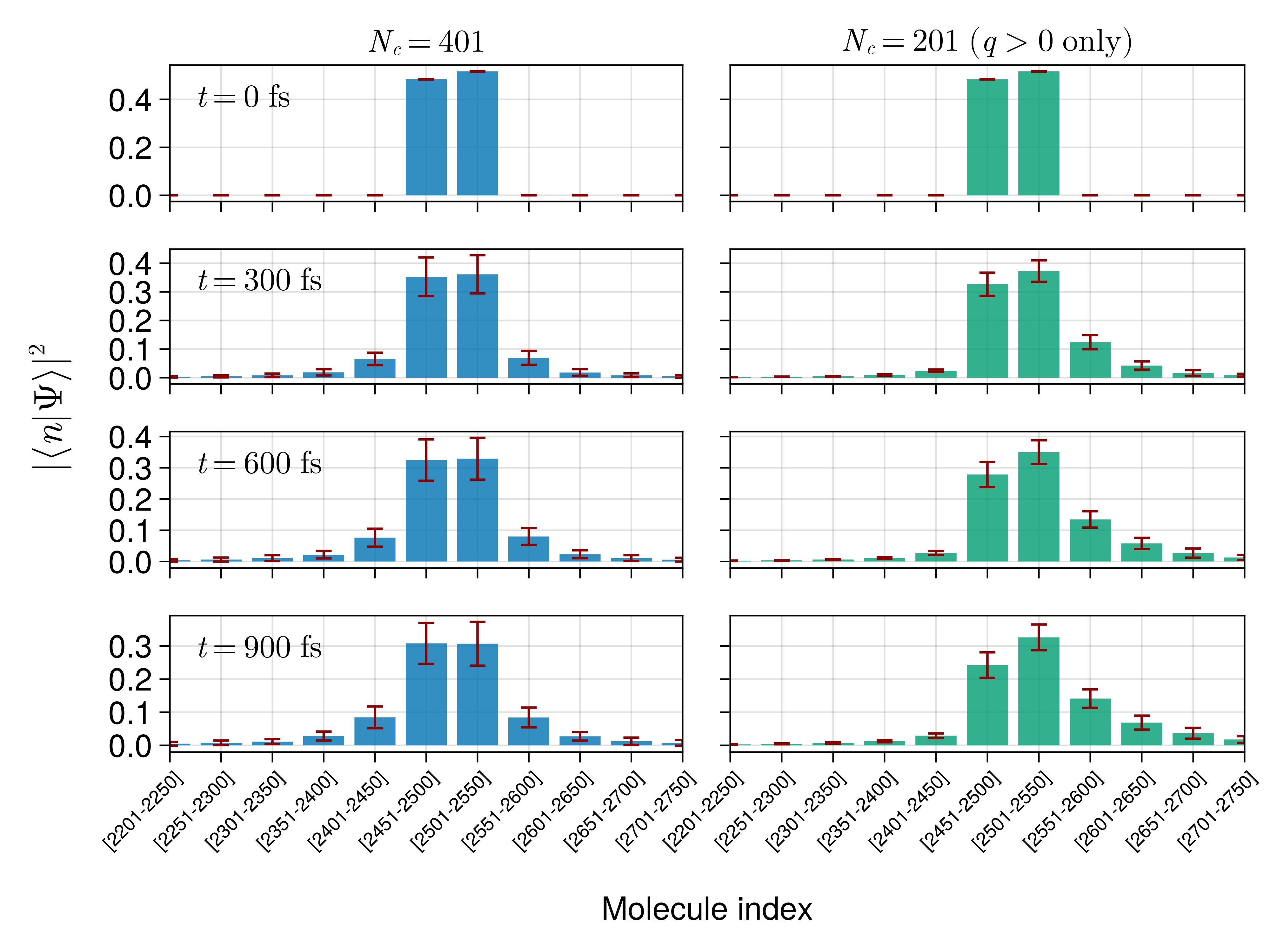}
    \caption{Wave packet grouped in bars over 50 molecules for a system with positive and negative (blue) and a system with only positive momentum photon modes (green). The number of molecules in this wire is 5000, and $\Omega_R = 0.1$ eV. The molecular parameters are $E_M = 2.0$ eV, $\sigma_M = 0.04$ eV, $a = 10$ nm, and $\sigma_a = 1$ nm. The initial wave packets have a width of 120 nm ($d(0) \approx 12$ molecules). Error bars show the standard deviation obtained from 100 realizations.}
    \label{fig:cav_momentum_dis}
\end{figure}

 Fig. \ref{fig:cav_momentum_dis} compares exciton wave packets obtained in simulations that retained only $q > 0$ cavity modes (right panel) to wave packets obtained in the presence of positive and negative quasimomentum modes (left panel) at different times in the presence of static disorder. The complete model has the wave packet spreading in both directions symmetrically. In contrast, the second model containing only $q>0$ shows a suppressed transport towards the left side of the wire. Since the energy transport in our model is solely mediated by photons, this result can be readily interpreted to arise from the lack of left-moving photons in the simulations that retained only $q > 0$ modes. 

\par We conclude that while an effective model that excludes photon modes traveling along a specific direction may be useful for the investigation of non-reciprocal systems with unidirectional energy transport, their predictions for coherent exciton mobility will likely be drastically overestimated in the presence of disorder due to the lack of backscattering processes. For instance, comparing the complete model (with doubly degenerate cavity modes) with the one-directional, we find that the probability of detecting an excited molecule with $x_n > x_0 + \sigma_x$ is roughly doubled for all time steps when photons are not allowed to have negative momentum. The situation becomes worse at later times due to the much weaker localization occurring in the model with no negative momentum modes.

\subsection{Time-resolved exciton localization}

\par
\begin{figure}
    \centering
    \includegraphics[width=\textwidth]{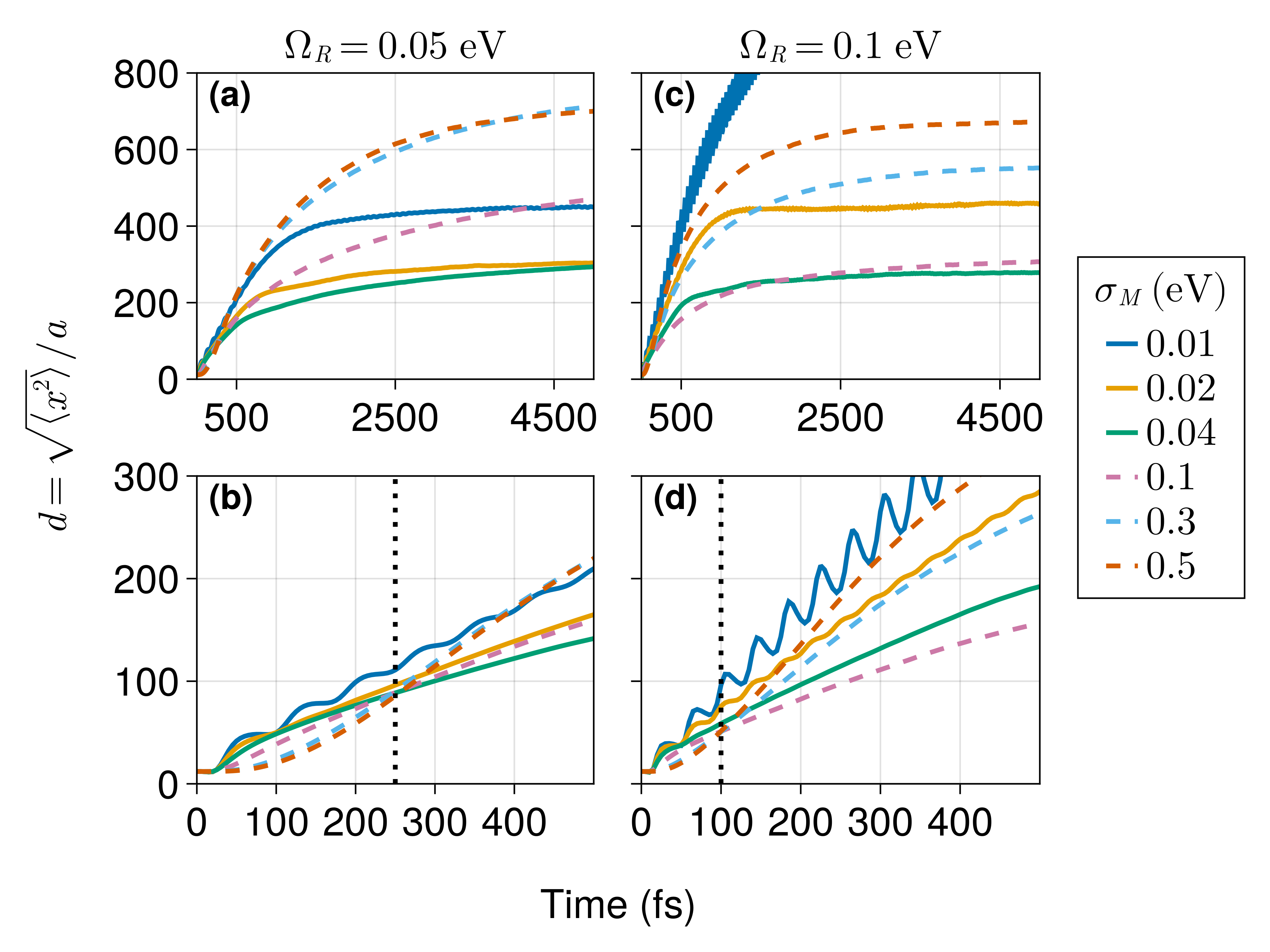}
    \caption{Wave packet spread ($d$) over time for several values of energetic disorder. The Rabi splitting is set to 0.05 eV (a and b), and 0.1 eV (c and d), and the cavity is in resonance with the average molecular excitation energy. (b) and (d) present views into the sub-ps window of (c) and (d), respectively, with dotted vertical lines indicating the time scale for DET emergence. Dashed lines represent trajectories where DET is observed. System size is $N_M = 5000$ with $N_c = 1001$. Each trajectory represents an average of over 100 realizations.}
    \label{fig:disp_prop}
\end{figure}

 We finish this work with an illustration of the main exciton transport phenomena exhibited by our model. Figs. \ref{fig:disp_prop}(a) and (c) show representative trajectories for the exciton width under a wide range of disorder strengths with $\Omega_R = 0.05$ and $\Omega_R = 0.1$ eV, respectively. These trajectories show ballistic transport in the early sub-ps window followed by a transient diffusive dynamics. For $\Omega_R = 0.1$ eV [Fig. \ref{fig:disp_prop}(c)], wave packet localization takes place within the time range of 2 to 4 ps for $\sigma_M/\Omega_R > 0.1$. In contrast, when 
$\Omega_R = 0.05$ and $\sigma_M/\Omega_R < 0.8$ [Fig. \ref{fig:disp_prop}(a)], 
localization happens at later times, between 4 and 5 ps. However, at larger values of $\sigma_M$, localization occurs on a longer time scale $t > 5 \text{ps}$. This is likely caused by strong disorder effects, which are observed earlier for smaller values of $\Omega_R$. Indeed, we also observe clear signatures of recently reported disorder-enhanced transport (DET) phenomena. \cite{Chavez2021Disorder-EnhancedCavities, Allard2022Disorder-enhancedPhotons, Dubail2022LargeCavity, Engelhardt2022PolaritionMicrocavities}. For example, Fig. \ref{fig:disp_prop}(c) shows the wave packet localization length is an increasing function of $\sigma_M$ for all values of $\sigma_M \geq 0.1$ eV.

Our wave packet simulations allow us to gain direct insight into the transient DET dynamics. We find that at very short times $t < 100 ~\text{fs}$ [Fig. \ref{fig:disp_prop}(b)], with $\Omega_R = 0.05$ eV, the exciton transport under strong disorder is essentially suppressed (highly subdiffusive). Figs. \ref{fig:disp_prop}(b) and \ref{fig:disp_prop}(d) also mark the crossover of exciton width trajectories under strong over weak disorder and allow us to visualize the time scales required for DET manifestation. These results show that DET emerges earlier under stronger collective coupling, as demonstrated by the shift from 250 to 100 fs of the time required for DET to operate and the corresponding shortening of the initial subdiffusive dynamics as $\Omega_R$ increases from 0.05 to 0.1 eV. 

\par Our findings show that the coherent evolution of exciton wave packets in photonic wires under strong light-matter coupling leads to DET operating on realistic time scales. However, the relevance of DET in the presence of dephasing induced by dynamical disorder (e.g., interaction with a thermal bath) remains an open question to be addressed in future work.

\section{Conclusions}
Our work presents several key features of space-time-resolved exciton wave packet evolution in a lossless polaritonic wire relevant to future theoretical and experimental investigations of polariton chemistry. We have reported time-resolved exciton dynamics explicitly showing ballistic, diffusive, and subdiffusive polariton-assisted exciton transport in disordered wires under a variety of conditions of weak and strong energetic disorder. Our simulations enabled our unveiling of a short period of largely suppressed exciton propagation under strong disorder. This feature is Rabi splitting dependent and arises prior to the onset of disorder-enhanced transport regime in  ultrafast timescales.

The convergence of our simulations was investigated thoroughly with respect to parameters such as cavity length (number of molecules) and number of photon modes. We found that a small number of molecules can reproduce the very early dynamics of the system appropriately. Still, qualitatively incorrect results arise when the wave packet spreads over a length scale of the same order as the system size (the observed exciton wave packet dynamics was essentially independent of the density of photonic states). We have also demonstrated that a multimode description of the radiation field, covering a sufficiently large energy interval, is necessary to properly describe the system evolution, especially at early propagation times. This energy range can be estimated as $0.4 \text{eV} + 2\Omega_R$. For example, when $\Omega_R = 0.1$ eV and $\omega_M = 2.0$ eV, we find that a cutoff energy of 2.6 eV includes the most relevant cavity modes.

In the presence of static disorder, the photonic distributions become more intricate, but the convergence trends observed in the absence of disorder are maintained. Hence, the simpler zero-disorder simulations requiring no ensemble averaging may provide a reference for constructing an optimal set of photon modes. Investigating the weight of each EM mode to the overall dynamics, we found that \textit{both} light-matter resonances and the mean initial exciton momentum play a fundamental role in determining the dominant microcavity modes. However, upon a substantial increase in the Rabi splitting or static disorder, we observe a transition into a quasiergodic regime where the photon weight approaches a uniform distribution over a large interval of wave vectors.

Overall, our findings highlight the rich diversity of exciton coherent transport phenomena in polaritonic wires and emphasize that a multimode description of the radiation field is essential to describe them accurately.
\begin{acknowledgement}
R.F.R. acknowledges generous start-up funds from the Emory University Department of Chemistry. 
\end{acknowledgement}

\bibliography{references}

\end{document}